\documentclass[aps,a4paper,amsmath,amssymb]{revtex4}
\usepackage{exscale,latexsym}
\usepackage[dvips]{graphicx}
\begin{document}
\preprint{KLT-\today}
\title{String theory and the KLT-relations between gravity and gauge theory including external
matter}
\author{N.~E.~J.~Bjerrum-Bohr$^{a,b}$ and K.~Risager$^a$}
\email{bjbohr@nbi.dk,risager@nbi.dk}
\affiliation{$^a$The Niels Bohr Institute,\\
Blegdamsvej 17, DK-2100 Copenhagen, Denmark\\ and \\
$^b$Department of Physics,\\
University of Wales, Singleton Park, Swansea, SA2 8PP, UK}
\date{\today}
\begin{abstract}
We consider the Kawai-Lewellen-Tye (KLT) factorizations of gravity scalar-leg amplitudes
into products of Yang-Mills scalar-leg amplitudes. We check and examine the factorizations
at $\mathcal O(1)$ in $\alpha'$ and extend the analysis by considering KLT-mapping in the case
of generic effective Lagrangians for Yang-Mills theory and gravity.
\end{abstract}\maketitle
\section{Introduction}
Recently, the so-called KLT-relations were reanalyzed from the perspective of effective field
theory~\cite{Bjerrum-Bohr:2003vy,B5}. It was found that KLT-relationships were valid for
effective amplitudes, and that effective generalizations were possible.
Non-trivial diagrammatic relationships between the amplitudes were also discovered
presenting interesting links between the pure Yang-Mills amplitudes as well as between
gravity and Yang-Mills amplitudes. It was suggested that introducing external source fields
in the actions might be a way to further extend the KLT-mapping. This paper aims to achieve
this by including into the formalism a massless scalar field.
Investigations of KLT in the presence of matter have previously been carried out in
ref.~\cite{Bern4} although in a different setting.

The fundamental dualities of string theory, linking closed and
open strings, contain also the possibility of combining seemingly
completely uncorrelated field theories at low energies,
\emph{i.e.},~below the Planck scale. String theory will at such
scales {\it essentially} correspond to a particular version of an
effective field theory, described by a local effective action
where the massive particle modes are integrated out, see {\it
e.g.}, refs.~~\cite{Tsey1,Tsey2,Tsey3,
  Gross:1986iv,Callan:jb,Gross:1985fr,Hull:yi,Sannan:tz}. An example
of a field theory relationship induced by string theory dualities
is the Kawai-Lewellen-Tye (KLT) relations first discovered in
ref.~\cite{KLT}. The KLT-relationship represents a curious
connection between two dissimilar field theories, namely
perturbative Yang-Mills theory and gravity. Being both non-Abelian
gauge theories they possess some similarities in their
description, but in their dynamical behavior they are quite
different. For example, Yang-Mills theory is an asymptotically
free theory at high energies, while gravity is well-defined in the
infrared but ultraviolet troubled. For example, the fundamental
Yang-Mills action is renormalizable at $D=4$, while gravity is
well-known to be non-renormalizable in four dimensions, see, {\it
e.g.},~ refs.~\cite{Veltman,NON,matter}.

Effective field theory~\cite{Weinberg,Donoghue:dn} provides a way
to resolve the renormalization difficulties of gravity. Treated as an
effective field theory the renormalization problem of the gravitational action
no longer exists because all higher derivative terms possibly generated
by the loop diagrams are already included in the action. Thus by each loop-order
all divergent loop-terms can be absorbed into the coupling constants of the theory.
See refs.~\cite{B1,B2,DB,QED} for some explicit calculations involving
effective gravity. To treat the Yang-Mills action by effective means
is a possibility although not a necessity for renormalization reasons,
at $D\leq4$.

The KLT-relations at tree level are directly linking amplitudes
in Yang-Mills theory and gravity~\cite{Berends}. Hence
useful knowledge about tree gravity scattering amplitudes can be
extracted from the much simpler Yang-Mills tree amplitudes.
Through cuts of loop diagrams, KLT-techniques can be applied with great
success in loop calculations -- using propagator cuts and the unitarity
of the $S$-matrix. As examples of such investigations, see {\it
e.g.},~ refs.~\cite{ Bern1,Bern2,Bern11,Bern0,Dunbar}. One important
result shown this way was that $\mathcal N=8$ SUGRA is less divergent
in the ultraviolet than was previously believed to be the case. For a
good review see, ref.~\cite{Bern2b}.

Investigations of factorization of gravity vertices at the
Lagrangian level have been carried out in ref.~\cite{Bern3}.
In ref.~\cite{Siegel} rather mysterious factorizations of gravity
vertices have been investigated employing a particular formalism
involving vierbeins.

The structure of this paper will be as follows. We will first briefly discuss the
KLT-relationship, next we will consider the fundamental Lagrangians and the
amplitudes in gravity and Yang-Mills theory. Through a number of examples
we will directly demonstrate that the KLT-relationship works for amplitudes
involving scalar fields -- the most challenging example
being a 5-point amplitude mapping. Next the effective extension of
this is discussed, and we present to lowest order the effective
Lagrangians in Yang-Mills theory and in gravity including the adequate
terms for the massless scalar field. Through a few basic examples we
will show that the KLT-mapping should be possible for effective
amplitudes too. Finally we will summarize and discuss what have been
achieved.

Throughout this paper we will use the $(+ - - -)$ metric and employ
the notation $s_{12}=2(k_1\cdot k_2), \ldots$. Helicity representations of amplitudes are employed
whenever useful.\footnote{\footnotesize{The helicity representation and
its twistor-space~\cite{Twister} connection have lately caught
attention initiated by the papers~\cite{Witten,Witten2}. The
viewpoint is essentially that perturbative gauge theory though the
helicity representation can be seen as a string theory in twistor-space.
This might also have interesting implications for the KLT-relationship.}}
We put $g=\frac{1}{2}$ and $\kappa=1$ in all calculations.

\section{Theory}
From string theory~\cite{GSW} we know that the generic $M$-point
amplitude of a closed string relates to a product of open string
amplitudes in the following way:
\begin{equation}\begin{split}
\label{eq:kltunreduced}
{{\cal A}}^M_{\rm closed} \sim \sum_{\Pi,\tilde \Pi}
e^{i\pi\Phi(\Pi,\tilde\Pi)}{{\cal A}}_M^\text{left open}(\Pi)
{\tilde{\cal A}}_M^\text{right open}(\tilde \Pi),
\end{split}\end{equation}
here $\Pi$ and $\tilde\Pi$ are the cyclic orderings associated with
the external open-string right and left moving sources.
The phase factor $\Phi(\Pi,\tilde \Pi)$ of the exponential
relates explicitly to the appropriate cyclic permutations of
the open string sources.

We have the following KLT-relations for the 3-, 4- and 5-point amplitudes:
\begin{equation}\begin{split}
{{\cal M}}_{\rm 3 \
gravity}^{\mu\tilde\mu\nu\tilde\nu\rho\tilde\rho}&(1,2,3) =
-i{{\cal A}}^{\mu\nu\rho}_\text{3 L-gauge}(1,2,3)\times
\tilde{\cal A}^{\tilde\mu\tilde\nu\tilde\rho}_\text{3
R-gauge}(1,2,3),
\end{split}\end{equation}

\begin{equation}\begin{split}
{\cal M}_{\rm 4 \
gravity}^{\mu\tilde\mu\nu\tilde\nu\rho\tilde\rho\sigma\tilde\sigma}&(1,2,3,4)
= -\frac{i}{\pi\alpha'}\sin(\pi s_{12}\alpha')\Big[ {\cal
A}^{\mu\nu\rho\sigma}_\text{4 L-gauge}(1,2,3,4)\times\tilde{\cal
A}^{\tilde\mu\tilde\nu\tilde\rho\tilde\sigma}_\text{4
R-gauge}(1,2,4,3)\Big],
\end{split}\end{equation}
and
\begin{equation}\begin{split}
    {\cal M}_{\rm 5 \
      gravity}^{\mu\tilde\mu\nu\tilde\nu\rho\tilde\rho\sigma\tilde\sigma\tau\tilde\tau}(1,2,3,4,5)
    &= -\frac{i}{\pi^2\alpha'^2} \Big(\sin(\pi s_{12}\alpha')\sin(\pi
    s_{34}\alpha')\Big[{\cal A}^{\mu\nu\rho\sigma\tau}_\text{5
      L-gauge}(1,2,3,4,5)\times\tilde{\cal
      A}^{\tilde\mu\tilde\nu\tilde\rho\tilde\sigma\tilde\tau}_\text{5
      R-gauge}(2,1,4,3,5)\Big]\\&+\sin(\pi s_{13}\alpha')\sin(\pi
    s_{24}\alpha')\Big[{\cal A}^{\mu\nu\rho\sigma\tau}_\text{5
      L-gauge}(1,3,2,4,5)\times\tilde{\cal
      A}^{\tilde\mu\tilde\nu\tilde\rho\tilde\sigma\tilde\tau}_\text{5
      R-gauge}(3,1,4,2,5)\Big]\Big).
\end{split}\end{equation}
In the above equations ${\cal M}$ is the gravity tree amplitude, and
${\cal A}$ is the color-ordered amplitude for the gauge theory. We
assume identical left and right-moving theories.  The explicit forms
of the above KLT-relations are fitted to match our conventions, which
differ from those of~\cite{KLT}.

To order $\mathcal O(1)$ in $\alpha'$, the above relations are reduced
in the following way:
\begin{equation}\begin{split}\label{eq22}
{{\cal M}}_{\rm 3 \
gravity}^{\mu\tilde\mu\nu\tilde\nu\rho\tilde\rho}&(1,2,3) =
-i{{\cal A}}^{\mu\nu\rho}_\text{3 L-gauge}(1,2,3)
\tilde{\cal A}^{\tilde\mu\tilde\nu\tilde\rho}_\text{3
R-gauge}(1,2,3),
\end{split}\end{equation}
\begin{equation}\begin{split}\label{eq23a}
{\cal M}_{\rm 4 \
gravity}^{\mu\tilde\mu\nu\tilde\nu\rho\tilde\rho\sigma\tilde\sigma}&(1,2,3,4)
= -is_{12}{\cal
A}^{\mu\nu\rho\sigma}_\text{4 L-gauge}(1,2,3,4)\tilde{\cal
A}^{\tilde\mu\tilde\nu\tilde\rho\tilde\sigma}_\text{4
R-gauge}(1,2,4,3),
\end{split}\end{equation}
and
\begin{equation}\begin{split}\label{eq23b}
    {\cal M}_{\rm 5 \
      gravity}^{\mu\tilde\mu\nu\tilde\nu\rho\tilde\rho\sigma\tilde\sigma\tau\tilde\tau}(1,2,3,4,5)
    =& -is_{12}s_{34}{\cal A}^{\mu\nu\rho\sigma\tau}_\text{5
      L-gauge}(1,2,3,4,5)\tilde{\cal
      A}^{\tilde\mu\tilde\nu\tilde\rho\tilde\sigma\tilde\tau}_\text{5
      R-gauge}(2,1,4,3,5)\\&-is_{13}s_{24}{\cal
      A}^{\mu\nu\rho\sigma\tau}_\text{5 L-gauge}(1,3,2,4,5)\tilde{\cal
      A}^{\tilde\mu\tilde\nu\tilde\rho\tilde\sigma\tilde\tau}_\text{5
      R-gauge}(3,1,4,2,5).
\end{split}\end{equation}

The fundamental Einstein-Hilbert Lagrangian in gravity, including a
real scalar field $\phi$, reads:
\begin{equation}
{\cal L} = \sqrt{-g}\Big[\frac{2R}{\kappa^2} + \frac12
g^{\mu\nu}D_\mu \phi D_\nu \phi\Big],
\end{equation}
where $\kappa^2=32\pi G$, $g_{\mu\nu}$ is the metric field, $g=\det
(g_{\mu\nu})$ and $R$ is the scalar curvature. Similarly in Yang-Mills
theory the corresponding fundamental Lagrangian is:
\begin{equation}
{\cal L} = {\rm tr}\Big[\frac{1}{4}{F}^2_{\mu\nu}-\frac12D_\mu
\phi D^\mu \phi\Big],
\end{equation}
where $A_\mu$ is the vector field, $\phi$ is a real field and in the adjoint
representation, and:
\begin{equation}
F_{\mu\nu} = \partial_\mu A_\nu - \partial_\nu A_\mu -
g[A_\mu,A_\nu].
\end{equation}
The trace is over the generators of the non-Abelian Lie-algebra. We
normalize such that ${\rm tr}(t^at^b)=-\delta^{ab}$ and
$[t^a,t^b]=-f^{abc}t^c$.

Treating gravity and gauge theory as effective theories, the Lagrangians of
both theories have to be augmented with all possible higher derivative terms.
For each of these terms we associate a factor of $(\alpha')^P$ where the power $P$ is governed
by the number of derivatives in the gravitational terms and by the mass dimension of
the gauge terms. Terms not invariant under field redefinitions will not alter
the S-matrix -- thus such terms can be neglected in Lagrangians.

\section{Results}
\subsection{The KLT-relations to leading order in $\alpha'$}
To verify the KLT-relations between the basic amplitudes we will
consider three types of amplitude mappings, all of which involve
scalar legs. We let $s$, $v$ and $h$ represent scalars, vectors
and gravitons, respectively.

Using the above Lagrangians we extract the Feynman vertex rules and
calculate the scattering amplitudes. We consider first the four scalar
leg amplitude. Here the basic one-graviton exchange diagram is related
to the one-vector exchange diagram through KLT.

On the gravity side we have:
\begin{equation}
{\cal M}(s_1,s_2,s_3,s_4) = \frac{i}{4}
\frac{s_{12}^2s_{13}^2+s_{13}^2s_{14}^2+s_{14}^2s_{12}^2}{s_{12}s_{13}s_{14}}=\frac{i}{16}\frac{(s_{12}^2+s_{13}^2+s_{14}^2)^2}{s_{12}s_{13}s_{14}},
\end{equation}
by the Mandelstam identity, while on the Yang-Mills side we have:
\begin{equation}
{\cal A}(s_1,s_2,s_3,s_4) = -i\frac14\frac{s_{12}^2+s_{13}^2+s_{14}^2}{s_{12}s_{14}}, \ \ \ \ {\rm and} \ \ \ \
\tilde{\cal A}(s_1,s_2,s_4,s_3) = -i\frac14\frac{s_{12}^2+s_{13}^2+s_{14}^2}{s_{12}s_{13}},
\end{equation}
such that:
\begin{equation}\begin{split}
-is_{12}{\cal A}(s_1,s_2,s_3,s_4)\tilde{\cal A}(s_1,s_2,s_4,s_3)& =
\frac {i
s_{12}}{16}\Big(\frac{s_{12}^2+s_{13}^2+s_{14}^2}{s_{12}s_{14}}\Big)\times
\Big(\frac{s_{12}^2+s_{13}^2+s_{14}^2}{s_{12}s_{13}}\Big)\\
&=\frac{i}{16}\frac{(s_{12}^2+s_{13}^2+s_{14}^2)^2}{s_{12}s_{13}s_{14}}={\cal
  M}(s_1,s_2,s_3,s_4),
\end{split}\end{equation}
It is seen that the KLT-relation precisely maps the product of the
left/right gauge theory amplitudes into the gravity amplitude.

Next we consider the KLT-mapping of a mixed amplitude -- the one we
will look at is the 2-scalar-2-graviton mapping into the product of
two 2-scalar-2-vector amplitudes. Only the partial amplitudes are used
where polarizations are contracted with polarizations and momentum
with momentum, and for clarity we have chosen to omit the products of
the polarizations vectors.  The diagrams which need to be considered
involve both contact terms and interaction terms. On the gravity side
one gets:
\begin{equation}
{\cal M}(s_1,s_2,h_3,h_4) = \frac{i}{4}\frac{s_{13}s_{14}}{s_{12}}.
\end{equation}
In a Yang-Mills theory with a mixed matter content there will be different
amplitude expressions depending on how we number the
particles and these amplitude expressions will relate independently
to gravity through KLT.
The KLT-relation for the above mixed process thus has two independent
forms and we hence need expressions for the following three amplitudes:
\begin{equation}
{\cal A}(s_1,s_2,v_3,v_4) = -\frac i2 \frac{s_{13}}{s_{12}},\quad
{\cal A}(s_1,s_2,v_4,v_3)=-\frac i2\frac{s_{14}}{s_{12}},\quad
{\rm and}\quad{\cal A}(s_1,v_3,s_2,v_4)=-\frac i2.
\end{equation}
By insertion we immediately see that the two independent gauge
amplitude products of the relation are equal to the gravity amplitude:
\begin{equation}
\label{eq:ssgg}
-is_{12}{\cal A}(s_1,s_2,v_3,v_4)
\tilde{\cal A}(s_1,s_2,v_4,v_3)
=-is_{23}{\cal A}(s_1,s_2,v_3,v_4)
\tilde{\cal A}(s_1,v_3,s_2,v_4)
= {\cal M}(s_1,s_2,h_3,h_4).
\end{equation}

Finally, we examine the factorization of the 4-scalar-1-graviton
amplitude into the product of two 4-scalar-1-vector amplitudes. This
example is the most involved of our checks and is a non-trivial check
of KLT. The verification relies on the helicity state notation, see~\cite{dixon} for
a nice review. We use the same conventions here. Below we present the
results for the case of a helicity (+) graviton or gluon.
On the gravity side we have, summing all diagrams which go into this
amplitude, that:
\begin{equation}
{\cal M}(s_1,s_2,s_3,s_4,h_5^+)= (\langle 13 \rangle \langle 42
\rangle [12][34]-\langle 12 \rangle \langle 34 \rangle
[13][42])\frac{(\langle 12 \rangle \langle 34 \rangle
\langle 13 \rangle \langle 42 \rangle + \langle 13\rangle \langle 42
\rangle \langle 14 \rangle \langle 23 \rangle + \langle 14 \rangle
\langle 23 \rangle \langle 12
\rangle \langle 34 \rangle)^2 }{8\prod_{i<j}\langle ij \rangle },
\end{equation}
where the identity
$
\langle 13 \rangle \langle 42 \rangle [12][34]-\langle 12 \rangle
\langle 34 \rangle [13][42]=
4i\varepsilon_{\mu\nu\rho\sigma}k_1^\mu k_2^\nu k_3^\rho k_4^\sigma
$
can be used.

The Yang-Mills side reads:
\begin{equation}
{\cal A}(s_1,s_2,s_3,s_4,v_5^+) = \frac{\langle 12 \rangle
\langle 34 \rangle \langle 13 \rangle \langle 42 \rangle +
\langle 13\rangle \langle 42 \rangle \langle 14 \rangle
\langle 23 \rangle + \langle 14\rangle \langle 23 \rangle
\langle 12 \rangle \langle 34 \rangle}{\sqrt{8}\langle 12 \rangle
\langle 23 \rangle \langle 34 \rangle \langle 45 \rangle\langle 51 \rangle}.
\end{equation}
These amplitudes satisfy the to independent KLT-relations for the
5-point functions:
\begin{equation}\begin{split}
{\cal M}(s_1,s_2,s_3,s_4,h_5^+)& = -i s_{12}s_{34} {\cal A}(s_1,s_2,s_3,s_4,v_5^+)
\tilde{\cal A}(s_2,s_1,s_4,s_3,v_5^+)
-is_{13}s_{24}{\cal A}(s_1,s_3,s_2,s_4,v_5^+)\tilde{\cal A}(s_3,s_1,s_4,s_2,v_5^+)\\
&= -i s_{12}s_{35} {\cal A}(s_4,s_1,s_2,s_3,v_5^+)
\tilde{\cal A}(s_3,s_4,s_2,s_1,v_5^+)
-is_{13}s_{25}{\cal A}(s_4,s_1,s_3,s_2,v_5^+)\tilde{\cal A}(s_2,s_4,s_3,s_1,v_5^+)
\end{split}\end{equation}
as can be seen by insertion (and a fair amount of algebra).

This concludes our set of examples of KLT-mapping to order $\mathcal O(1)$ in $\alpha'$
in the presence of matter.

\subsection{Effective extensions to $\mathcal O(\alpha')$}
We now consider the KLT-relations in the case of the
effective extension of the two theories. The most general
gravitational Lagrangian to $\mathcal O(\alpha')$ can be written as:
\begin{equation}\begin{split}
    {\cal L} &= \sqrt{-g}\Big[\frac{2R}{\kappa^2} + \frac12
    g^{\mu\nu}D_\mu \phi D_\nu \phi+\alpha'
    \Big(c_1\kappa^{-2} R_{\mu\nu\alpha\beta}^2 +
    c_2\kappa^{-2}R_{\mu\nu}^2 + c_3\kappa^{-2}R^2 +
    c_4\kappa^2D_\mu \phi D^\mu \phi D_\nu \phi D^\nu \phi \\&+
    c_5\kappa^2\phi D_\mu \phi D_\nu \phi D^\mu D^\nu \phi
    + c_6\kappa^2\phi^2 D_\mu D_\nu \phi D^\mu D^\nu
    \phi+ c_7\phi^2 R_{\mu\nu\alpha\beta}^2 +
    c_8\phi^2 R_{\mu\nu}^2 + c_9\phi^2 R^2 +
    \ldots \Big) \Big],
\end{split}\end{equation}
where the ellipses denote terms higher order terms not necessary
for the present analysis.

By a field reparametrization of $\phi$ and $g_{\mu\nu}$, all coefficients
but $c_1$, $c_4$ and $c_7$ can be set to any desired value. The terms
$c_1$ and $c_7$ are left unchanged by such a reparametrization,
while $c_4$ picks up contributions from terms being altered under
the reparametrization. Thus, to generate on-shell (4-particle) amplitudes
we may limit ourselves to the effective Lagrangian:
\begin{equation}\begin{split}
{\cal L} &= \sqrt{-g}\Big[\frac{2R}{\kappa^2} + \frac12
g^{\mu\nu}D_\mu \phi D_\nu \phi +\alpha' \Big(c_1\kappa^{-2}
G_2+ c_4\kappa^2D_\mu \phi D^\mu \phi D_\nu \phi
D^\nu \phi +
c_7\phi^2 R_{\mu\nu\alpha\beta}^2\Big) \Big],
\end{split}\end{equation}
where $G_2=R_{\mu\nu\alpha\beta}^2-4R_{\mu\nu}^2+R^2$ is the four
dimensional Gauss-Bonnet invariant. This effective extension produces
effective corrections to all pure graviton vertices, corrections to
all vertices containing four scalars, and a correction to the
2-scalar-2-graviton vertex.\footnote{\footnotesize{We use $G_2$ instead of,
\emph{e.g.},~$R_{\mu\nu\alpha\beta}^2$ since that avoids the presence
of an effective correction to the graviton propagator in any
dimension.}}

In the Yang-Mills theory we have a similar situation. We have to include
all operators that contain one factor of $\alpha'$, \emph{i.e.},~
those of mass dimension six:
\begin{equation}\begin{split}
{\cal L}& = {\rm tr}\Big[\frac{1}{4}{F}^2_{\mu\nu}-\frac12D_\mu
\phi D^\mu \phi+ \alpha'\Big(a_1 (D_\mu F_{\mu\nu})^2 +
a_2gF_{\mu\nu} [F_{\nu\lambda}, F_{\lambda\mu}]
+a_3g[\phi,D_\mu\phi]D_\nu F_{\mu\nu}\\&
+ a_4g^2[\phi ,F_{\mu\nu}][\phi ,F_{\mu\nu}]
+a_5g^2[\phi,D_\mu\phi][\phi,D_\mu\phi]
+a_6g^2\phi D_\mu D_\mu D_\nu D_\nu \phi\Big)\Big].
\end{split}\end{equation}
By a field reparametrization we may set $a_1$, $a_3$ and $a_6$
to zero while allowing a change in the coefficients $a_2$, $a_4$
and $a_5$. Doing this we end up with the result:
\begin{equation}\begin{split}
{\cal L}& = {\rm tr}\Big[\frac{1}{4}{F}^2_{\mu\nu}-\frac12D_\mu
\phi D^\mu \phi+ \alpha'\Big(
a_2gF_{\mu\nu} [F_{\nu\lambda}, F_{\lambda\mu}]
+ a_4g^2[\phi ,F_{\mu\nu}][\phi ,F_{\mu\nu}]
+a_5g^2[\phi,D_\mu\phi][\phi,D_\mu\phi]\Big)\Big].
\end{split}\end{equation}
This effective extension produces corrections to the three-vector
vertex, the 2-scalar-2-vector vertex and the four scalar vertex, as
well as corrections to vertices not used in this context.

\subsection{Matching of effective operators}
The consequences of these extensions are next explored. The KLT-relations
hold in string theory order by order in $\mathcal O(\alpha')$ and it is the
expectation from previous investigations ref.~\cite{Bjerrum-Bohr:2003vy,B5}
that one should expect an effective mapping between the effective field theory
operators we have included in the effective Lagrangians. In order to
examine this, we have looked into two examples of such possible
mappings; namely the effective extensions of the four scalar
amplitudes and the 2-scalar-2-graviton/2-scalar-2-vector amplitudes.
Both these amplitude mappings will involve effective operators at
order $\mathcal O(\alpha')$ and hence relate the effective operators of gravity
to those of Yang-Mills through the 4-point KLT-relation.

In the case of the four scalar amplitude we have generated the following
effective field theory amplitude to order $\mathcal O(\alpha')$.

On the gravity side we have:
\begin{equation}\begin{split}
    {\cal M}(s_1,s_2,s_3,s_4) &= \frac i
    {16}\Big[\Big(\frac{(s_{12}^2+
    s_{13}^2+s_{14}^2)^2}{s_{12}s_{13}s_{14}}\Big)
    +32\alpha'c_4(s_{12}^2+s_{13}^2+s_{14}^2)\Big],
\end{split}\end{equation}
while on the Yang-Mills side we have:
\begin{equation}
{\cal A}(s_1,s_2,s_3,s_4) = \frac i4\frac{-(s_{12}^2+s_{13}^2+s_{14}^2)
-6a_5\alpha's_{12}s_{13}s_{14}}{s_{12}s_{14}}.
\end{equation}
Matching these amplitudes through KLT at order $\mathcal O(\alpha')$ leads to
the coefficient relationship:
\begin{equation}
c_4=\frac38a_5,
\end{equation}
which has to hold in order for the mapping to take place.

Looking at the effective amplitudes in the
2-scalar-2-graviton/2-scalar-2-vector case we have on the gravity
side:
\begin{equation}\begin{split}
{\cal M}(s_1,s_2,h_3,h_4) &= i\Big[\frac{s_{13}s_{14}}{4s_{12}}+
\alpha'\Big(-\frac{c_1}{4}s_{13}s_{14}+c_7s_{12}^2\Big)\Big],
\end{split}\end{equation}
while the necessary Yang-Mills amplitudes are:
\begin{equation}\begin{split}
    {\cal A}(s_1,s_2,v_3,v_4)=&
    i\Big[-\frac{s_{13}}{2s_{12}}+\alpha'\Big(-\frac12 a_4
    s_{12}+\frac{3}{4}a_2 (s_{13}-s_{14})\Big)\Big],\\
    {\cal A}(s_1,s_2,v_4,v_3)=&
    i\Big[-\frac{s_{14}}{2s_{12}}+\alpha'\Big(-\frac12 a_4
    s_{12}+\frac{3}{4}a_2 (s_{14}-s_{13})\Big)\Big],\\
    {\cal A}(s_1,v_3,s_2,v_4)=&
    i\Big[-\frac{1}{2}+\alpha'\Big(- a_4 s_{12}\Big)\Big].
\end{split}\end{equation}
Relating these equations in the same way as in (\ref{eq:ssgg}), we
derive the constraints:
\begin{equation}
c_7=0\qquad{\rm and}\qquad c_1=6a_2 \qquad{\rm and}\qquad a_4=-\frac32a_2.
\end{equation}
The equation $c_1=6a_2$ is noted to be exactly what was found in
ref.~\cite{{Bjerrum-Bohr:2003vy},B5} relating the pure effective
3-amplitude on the gravity side to the 3-amplitude on the Yang-Mills
side. The relationship $a_4=-\frac32a_2$ originates from the requirement
that the two 'gauge sides' of the KLT-relation must be equal. It
should be noted that, while the coefficient equations hold to all
orders of $\alpha'$, we cannot from the present analysis determine,
if at $\alpha'^2$, new coefficient equations generated from the
amplitude mapping will constrain the above $\mathcal O(\alpha')$ mapping
relationships.

\section{Discussion}
We have directly shown that the KLT-relations also hold for a number
of amplitudes involving external scalar legs, and that in such cases,
the factorization of gravity amplitudes into Yang-Mills amplitudes is
possible.  We have also shown through some preliminary examples that
an effective extension of these results seems possible, and that this
might be a key to gain more insight in the mapping of gravity
effective field theory operators into effective Yang-Mills operators.
The examples with the four scalar amplitudes and the
2-scalar-2-graviton/2-scalar-2-vector amplitudes clearly suggest that
we might gain some important insight including scalars into the
effective KLT-mapping, although the present examples should be
followed up by some heavier calculations involving more complicated
operators.

The amplitude factorization is possible to recast into direct use in
the process of calculating gravity amplitudes from much simpler
Yang-Mills amplitudes including matter fields.

It is not possible from our  present calculation to tell much about an
effective generalization of the mapping. In the paper ref.~\cite{B5}
we replaced the sine functions in the KLT-relations with arbitrary
polynomials. Such a generalization of the mapping involving matter
should be possible in the scalar approach as well, however one has to
go to next order in $\alpha'$, $\mathcal O(\alpha'^2)$, to get enough
information about how this extension should work.

The starting point for additional investigations in this field could
be to look into the effective KLT-mapping of the
4-scalar-1-graviton/4-scalar-1-vector amplitudes, since it
simultaneously relates all effective gravity and Yang-Mills operators
used here.

If the KLT-relations have a more fundamental meaning -- relating
gravity and Yang-Mills theory, the effective mapping of operators --
in or without the presence of matter seems to be an adequate starting
point for additional theoretical investigations.

\begin{acknowledgments}
  We would like to thank prof. Poul Henrik Damgaard and prof. David C.
  Dunbar for enlightening and good discussions. N.~E.~J.~Bjerrum-Bohr
  would like to thank the Leon Rosenfeld Scholarship Foundation for
  financial support.
\end{acknowledgments}

\end{document}